# Symmetry control of strong chiral light–matter interactions in photonic nanocavities for efficient circularly polarised emission.

*Rahul Kumar*[1], Ben Trodden[1], Anastasiia Klimash[1], Affar Karimullah[1], Nikolaj Gadegaard[2], Peter J. Skabara[1], Shun Hashiyada,[3, 4] Gordon J. Hedley[1*] and Malcolm Kadodwala[1]**

[1] School of Chemistry, Joseph Black Building, University of Glasgow, Glasgow, G12 8QQ, UK

[2] School of Engineering, Rankine Building, University of Glasgow, Glasgow G12 8LT, U.K

[3] Innovative Photon Manipulation Research Team, RIKEN Center for Advanced Photonics, 2-1 Hirosawa, Wako, Saitama 351-0198, Japan

[4] Department of Electrical, Electronic, and Communication Engineering, Chuo University, 1-13-27 Kasuga, Bunkyo-Ku, Tokyo 112-8551, Japan.

## Abstract

Chiral excited electronic states of molecules have an intrinsic sense of handedness, or twist, and are the active component in energy efficient display technologies and in new photosynthetic routes to produce pharmaceuticals. Creating chiral states is achieved by manipulating the "twistiness" of the geometric molecular structure. This is a demanding problem adding complexity due to the need to precisely control molecular geometry. Here we demonstrate a novel concept for creating chiral excited states which does not rely on molecular structure. Instead, it depends on hybridising a non-chiral molecule with a chiral electromagnetic field, producing a hybrid light-matter chiral polariton state. This is achieved by a symmetry-controlled strong chiral-light–matter interaction between an electromagnetic mode of a chiral nanocavity and an achiral molecule, a concept referred to as the electromagnetic-enantiomer. This electromagnetic mechanism simplifies the creation of chiral electronic states since it is far less demanding in terms of materials design. We have illustrated the concept using an exemplar system relevant to organic optoelectronic technology, producing efficient circularly polarised emission from a non-chiral emitter molecule.

# Introduction

Chirality, the property of asymmetry, is generally perceived as a geometric concept, with chiral materials existing in otherwise identical mirror forms known as enantiomers. Thus, creating chiral molecules, which have applications spanning pharmaceuticals to next generation photonic technologies, requires precise control of 3-D structure, a challenging chemical problem. However, chirality is not just a physical property, it is also possessed by electromagnetic (EM) fields. Circularly polarised (CP) light is chiral and has long been used as a "chiral influence" to induce asymmetry in chemical reactions[1-3]. Unfortunately, under normal conditions, CP light illumination cannot induce chirality in geometrically non-chiral (achiral) molecules, thus removing a potentially straightforward route to achieve chiral functionalities. In this work we demonstrate a simple EM chiral induction mechanism which is based on the strong coupling between the chiral near fields of an asymmetric photonic nanocavity and an achiral molecule placed within its vicinity, leading to the production of hybrid light-matter chiral polariton states (a polariton is a quasiparticle formed from the strong coupling of a photon with an exciton). We refer to molecules chirally perturbed in this way as EM-enantiomers, this concept is illustrated in Figure 1. The level of coupling between the chiral near fields and the molecule is dependent on the symmetries of both CP light and the chiral nanocavity, with symmetry match combinations leading to higher intensity fields and hence strong coupling. We exploit this symmetry-dependent strong coupling to create a composite polymer film that displays photoluminescence with a high level of circular polarisation (~ 30%). This level of performance has only previously been obtained from chiral molecular materials which have been complexly designed to display chirality on a macromolecular lengthscale[4-6]. This illustrates how the EM-enantiomer concept could be used to simplify the design and construction of organic light emitting diode (OLED) sources of CP light, which are required for energy efficient displays.

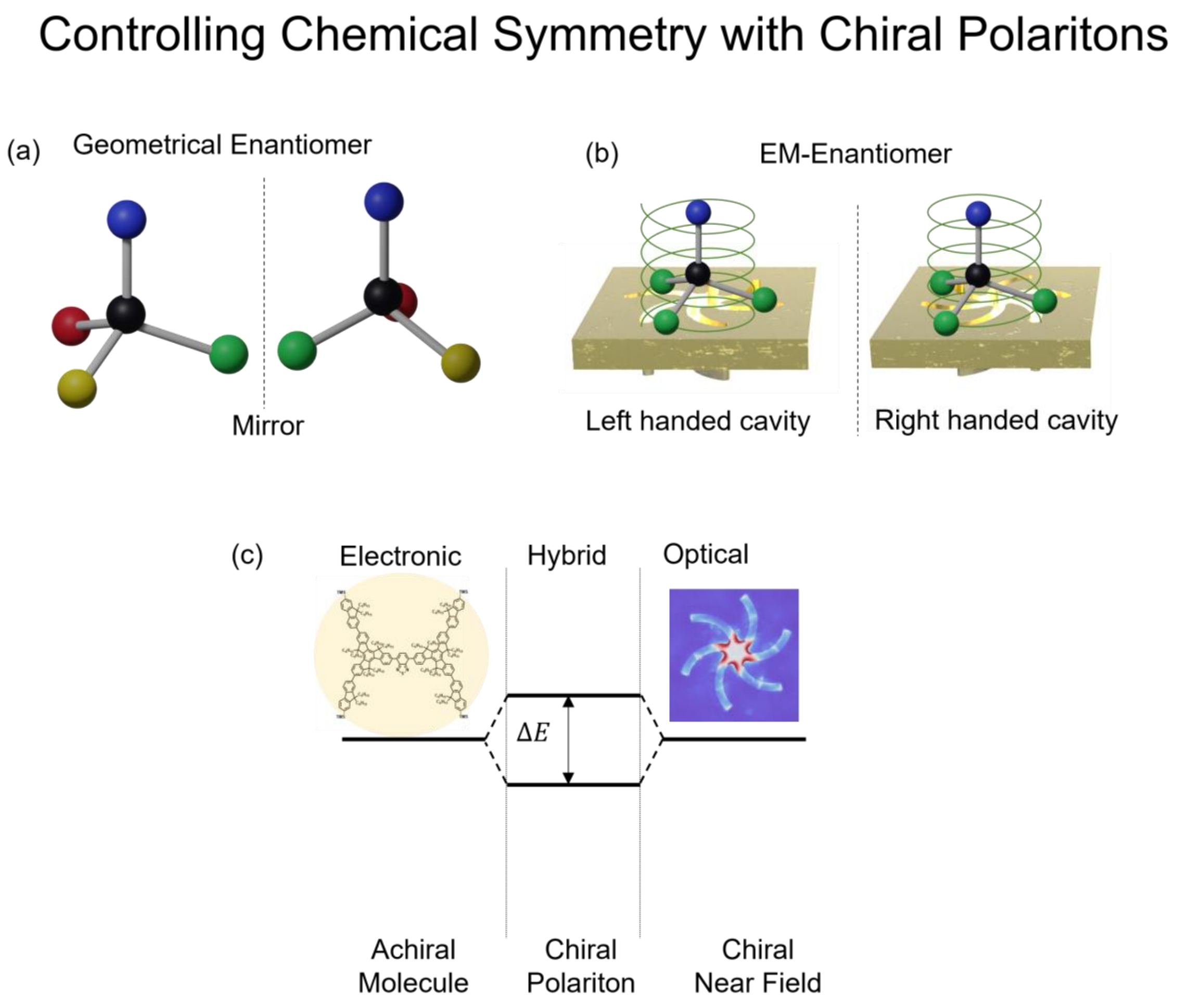

*Figure 1 The principles underpinning the electromagnetic enantiomer concept are illustrated. (a) A pair of enantiomers with mirror image geometries are shown. (b) The interaction of chiral nearfields with achiral molecules results in EM-enantiomers, through the symmetry dependent creation of polariton states. (c) A hybridisation diagram illustrating the splitting of states is shown*

There are two extremes in the level of light - matter interactions: weak coupling where the molecular excitation is unperturbed; and strong coupling where hybrid electronic and photonic (polaritonic) states are created. The formation of hybrid light-matter polaritonic states leads to the splitting of spectral bands, known as Rabi splitting[7]. Strong and weak levels of coupling between electronically excited states and plasmonic modes have been reported in a range of nonchiral systems (i.e. the EM coupling fields have no chiral asymmetries)[8-10]. The few studies of coupling of achiral molecular excitation to chiral modes report only weak coupling[11-13]. An experimental fingerprint of the strong coupling regime is the splitting of luminescence peaks, (> 50 meV) [8, 10, 14-17]. For achiral systems, strong coupling of excited molecular states to optical modes of cavities and antennae has been used to manipulate the selectivity of chemical reactions. This strategy is sometimes referred to as polaritonic chemistry and relies on the hybridisation of states modifying activation barriers and altering excited state lifetimes [14, 18-23], thus altering reaction pathways.

This study introduces the fascinating concept that strong chiral light-matter interactions can be used to manipulate molecular symmetry and hence be a tool in chiral (asymmetric) chemistry, a new paradigm of polaritonic chemistry.

# Results

We synthesised the molecule 4,7-bis(7,12-bis(9,9-dioctyl-7-(trimethylsilyl)-9H-fluoren-2-yl)-5,5,10,10,15,15-hexahexyl-10,15-dihydro-5H-diindeno[1,2-a:1',2'-c]fluoren-2-yl)benzo[c][1,2,5]thiadiazole (MHeB14), figure 2 (a), which is a candidate material for optoelectronic applications such as OLED technology. The experimental procedure and characterisation of the compound can be found in the supplementary information section. MHeB14 was designed for this work for its symmetrical structure, high photoluminescence quantum yield, photostability and compatibility in polymer blends, and is analogous to some star-shaped materials that have been studied by us for amplified spontaneous emission. Polymethyl methacrylate (PMMA) layers (~ 150 nm thickness), which were either undoped or doped with nominal 1, 10 and 50% w/w of MHeB14, were spin coated onto two types of nanostructured substrates. The substrates, which have been described in detail elsewhere[24], consist of ~ 100 nm thick Au films, deposited on to polycarbonate templates which contained periodic square arrays of either left-, or right-handed shuriken indentations. We have used two types of templated structures, they both contain identical shuriken indentations, but with periodicities of 1000 and 1500 nm respectively, figure 2(b). Subsequently, the two periodic nano-cavities will be referred to as 1000 nm and 1500 nm substrates. As will be shown later, although the shuriken structures are identical, these substrates have dissimilar EM environments. Effectively they allow the EM environment to be modified without altering the structure of the nanocavity, mitigating against possible influences of changing the nanocavity on the polymer film structure.

For reference purposes we have compared luminescence and absorption spectra for solutions of MHeB14 with the equivalent data from PMMA doped films deposited on unstructured Au surfaces. This reference data, figure 2(c), demonstrates that the optical absorption and emission properties of MHeB14 immobilised in the PMMA are not significantly modified compared to that of the free molecule in solution, figure 2(c), which is indicative of an absence of molecular aggregate formation. The slight red shift of the luminescence compared to solution can be attributed to different dielectric environments and the reduced conformational flexibility of MBeH14 in the matrix. The MBeH14 does not display significant optical absorption for wavelengths > 450 nm, which is the region occupied by resonances associated with Au $d \rightarrow sp$ inter-band transitions, figure 2(d), and plasmonic modes. Consequently, reflectance (linearly polarised) spectra collected from undoped and doped metafilms are only dominated by features associated with inter-band transitions (~500-680 nm) and plasmonic resonances (> 680 nm), see supplementary information.

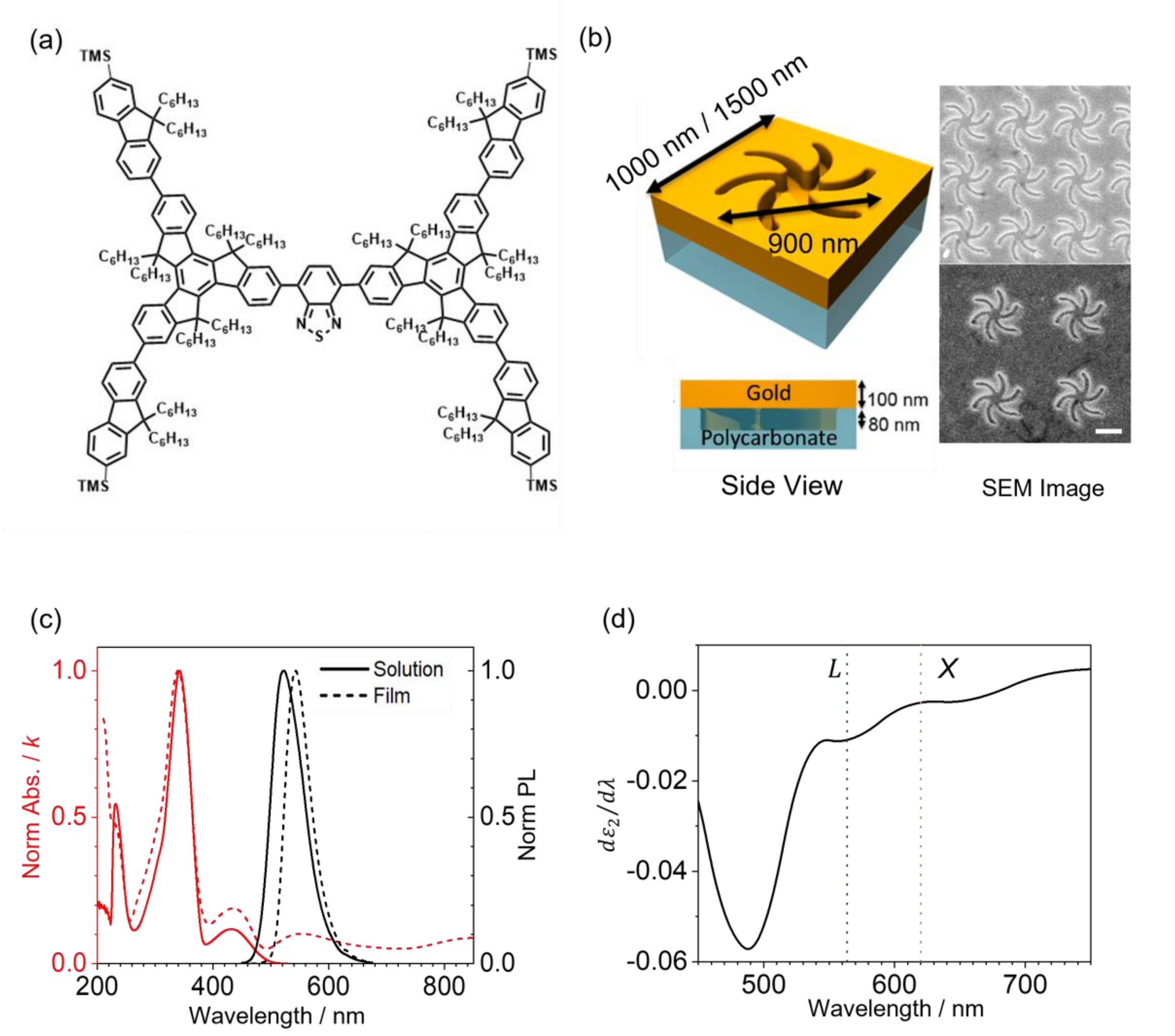

*Figure 2 (a) Molecular structure of MHeB14 dye. (b) Schematic showing orthographic (top) and side view (bottom) of LH nanocavity. (c) The self-normalised absorbance (solid red) and luminescence (solid black) of dye in toluene, imagery part of refractive index (dashed red), luminescence (dashed black) of 50% film, respectively. (d) Derivative of imaginary part of the permittivity ($\varepsilon_2$) of the flat gold layer obtained from ellipsometer showing two interband transitions L ($\lambda$ ~563 nm) & X ($\lambda$~620 nm).*

Circularly polarised luminescence spectra derived from detecting either left or right-handed light (LCP and RCP) are collected from both the undoped and doped metafilms. For clarity only the spectra for the 50% doping are discussed, other data can be found in supplementary information, but all dopant levels display qualitatively similar behaviour. The luminescence from the undoped films is derived from the radiative decay of holes in the 5*d* band, which produces very weak emission (two orders of magnitude weaker than from the 1% doped PMMA films) spanning the blue to IR regions of the spectrum [25]. The doped metafilms display a relatively larger amount of emission over the interband and plasmonic regions compared to the reference films on unstructured Au. A fraction of molecular luminescence from the doped metafilms displays helicity dependent structure, which is not observed in reference spectra in the absence of the nanocavities. In the case of doped 1000 nm metafilms, the luminescence showed two distinct regions of structure centred at ~ 560 and 620 nm, figure 3, which are regions associated with $d \rightarrow sp$ transitions (at the $L$ and $X$ symmetry points), with the most intense and obvious structure occurring at the $L$ region. For the $L$ region, the structure observed for mis-matched combinations of metafilm handedness and emitted light helicity (LH/RCP and RH/LCP) is a single peak, which appear to split into two less intense components for matched combinations, figure 4, with the splitting being equivalent to ~ 107 meV (Δλ~ 28 nm). Relatively weaker structure occurs in the region of the *X 5d-6sp* transition, with a single peak observed for the mis-matched combinations and weaker, less defined features observed for the matched combinations. For the 1500 nm metafilms, the structure is less pronounced and the most noticeable feature, a single peak, occurs close to the *X 5d-6sp* transition, while a weaker peak is observed in the vicinity of the *L 4d-6sp* transition. In contrast to 1000 nm periodicity, switching helicity does not result in the splitting of the peaks, but rather causes only a change in relative intensities of the structure. As with the 1000 nm case, the most intense structured luminescence is observed for the matched combinations of metafilm handedness and light helicity. For both periodicities the positions of the structured emissions do not significantly change with increasing level of doping. This is consistent with the structures being associated with inter-band transitions, which unlike plasmonic resonances, are relatively insensitive to the refractive index of the surrounding environment.

Extinction spectra derived from monitoring the scattering of LCP and RCP light display qualitatively similar behaviour to the circularly polarised luminescence data. It should be highlighted that the extinction spectra have more structured backgrounds than the equivalent luminescence data. For the 1000 nm metafilms splitting of peaks associated with the $L$ and $X$ $d \rightarrow sp$ transitions are discernible above the structured background for the matched symmetry combinations and not for the mis-matched ones.

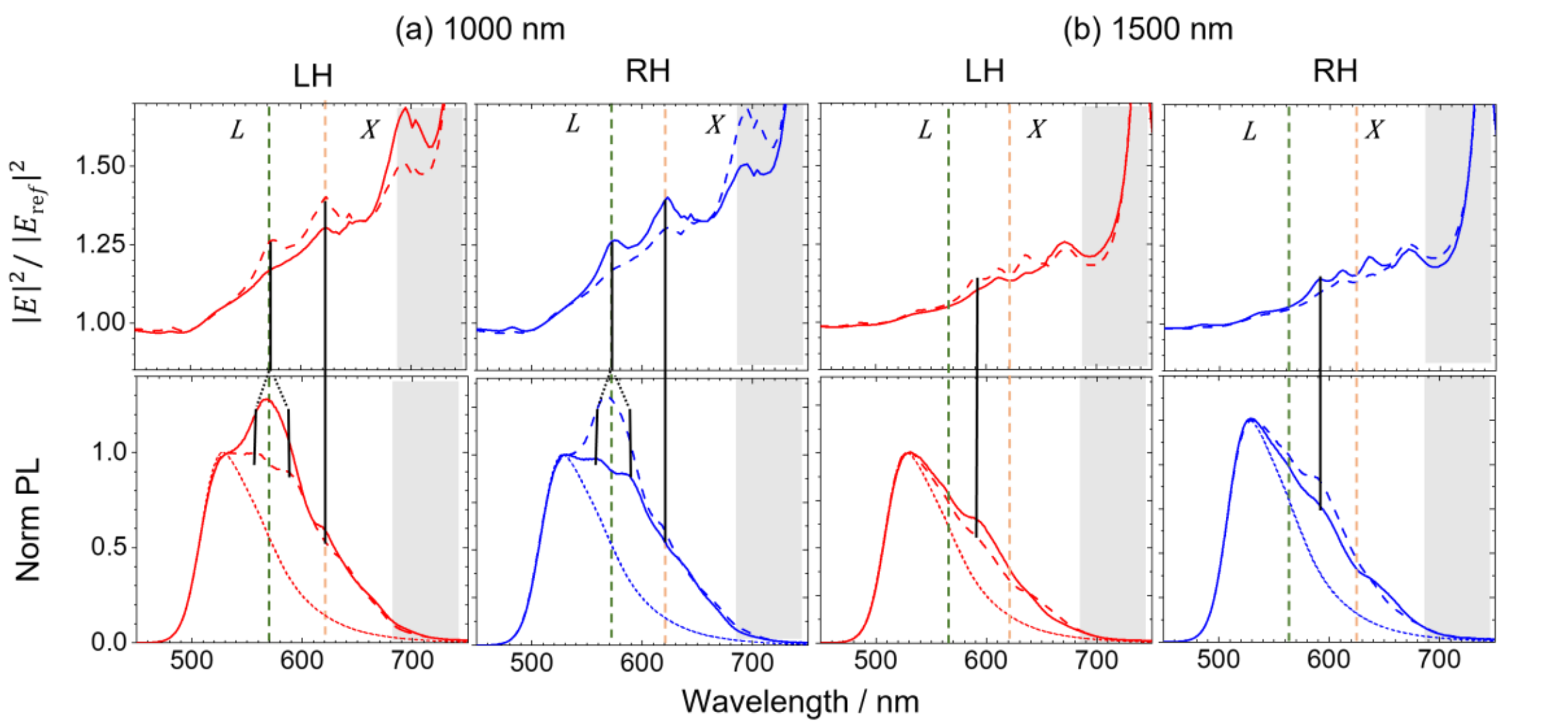

*Figure 3 The simulated relative electric-field intensity in 200 nm thick volume for 50% concentration dye (top) and normalised -PL (bottom) for (a) 1000 nm (b) 1500 nm metafilms. The black vertical lines are guidance for eyes, highlighting how the presence or absence of peak splitting is related to the intensities and positions of resonances in the EM fields. The dashed line and solid line respectively show the electric RCP and LCP PL emission from the structures. The overlapping dotted lines show LCP and RCP PL emission of dye when present on the flat gold. All spectra have been self-normalised to value at $\lambda$ =533 nm.*

The splitting of peaks in extinction spectra can arise through mechanisms other than strong coupling, e.g. Fano interference[26]. However, observing equivalent splitting in luminescence is a definitive fingerprint of strong coupling[16]. This is because PL is an incoherent process, in contrast to light scattering, and thus does not display Fano interference, hence splitting of a luminescence peak is a definitive signature of strong coupling[27]

Two significant inferences can be drawn from the experimental data about the nature of the EM environments within the cavities that define the strength of the coupling. The absence of strong coupling in the 1500 nm cavity indicates relatively weak electric fields compared to those for the 1000 nm. Most significantly, the intensities of the E fields in the 1000 nm cavity have a strong dependency on both cavity handedness and emission helicity. These phenomenological arguments can be justified by using EM numerical simulations to calculate the properties of the near fields, with the caveat that EM numerical simulation cannot fully describe excitation of inter-band transitions. This is because quantum mechanics is required to rationalise the transition between two electronic states and cannot be accounted for with classical Maxwell's equations [28, 29]. However, EM simulations can provide a guide to the nature of the near fields generated by light scattering, albeit they will overall underestimate the strength of the fields, since the effects of the inter-band transitions are not taken into consideration. Near field electric field enhancement ($|E|^2/|E_{ref}|^2$) for LCP and RCP light for both enantiomorphs of 1000 and 1500 nm metafilms have been calculated and compared to experimental luminescence spectra, figure 3, where $|E|$ and $|E_{ref}|$ are the volume averaged electric fields in the polymer films surrounding the nanocavities and an unstructured Au surface respectively. In addition, the level of chiral asymmetries of near fields have been parametrised with the optical chirality factor ($C$) [30, 31]. For the 1000 nm substrate there are peaks in $|E|^2/|E_{ref}|^2$ which coincide with the regions of enhanced structured emission. Consistent with the proposal that splitting of the luminescence structure is due to strong coupling, the largest peaks in $|E|^2/|E_{ref}|^2$ occur for matched combinations of light helicity and nanocavity handedness (*i.e.* LCP/LH and RCP/RH), figure 4. The greater intensities of the single luminescence bands observed for the weakly coupled mis-matched combinations can be attributed to the Purcell effect amplifying emission[32]. For the 1500 nm substrates the peaks in $|E|^2/|E_{ref}|^2$ also overlap with the peak of enhanced emission. However, the peaks in $|E|^2/|E_{ref}|^2$ are smaller than for the 1000 nm substrates, which is consistent with the weaker nature of the coupling, implied by the absence of peak splitting. Hence, for this weak coupling regime only the Purcell effect operates leading to enhanced emission coinciding with the peaks in the $|E|^2/|E_{ref}|^2$.

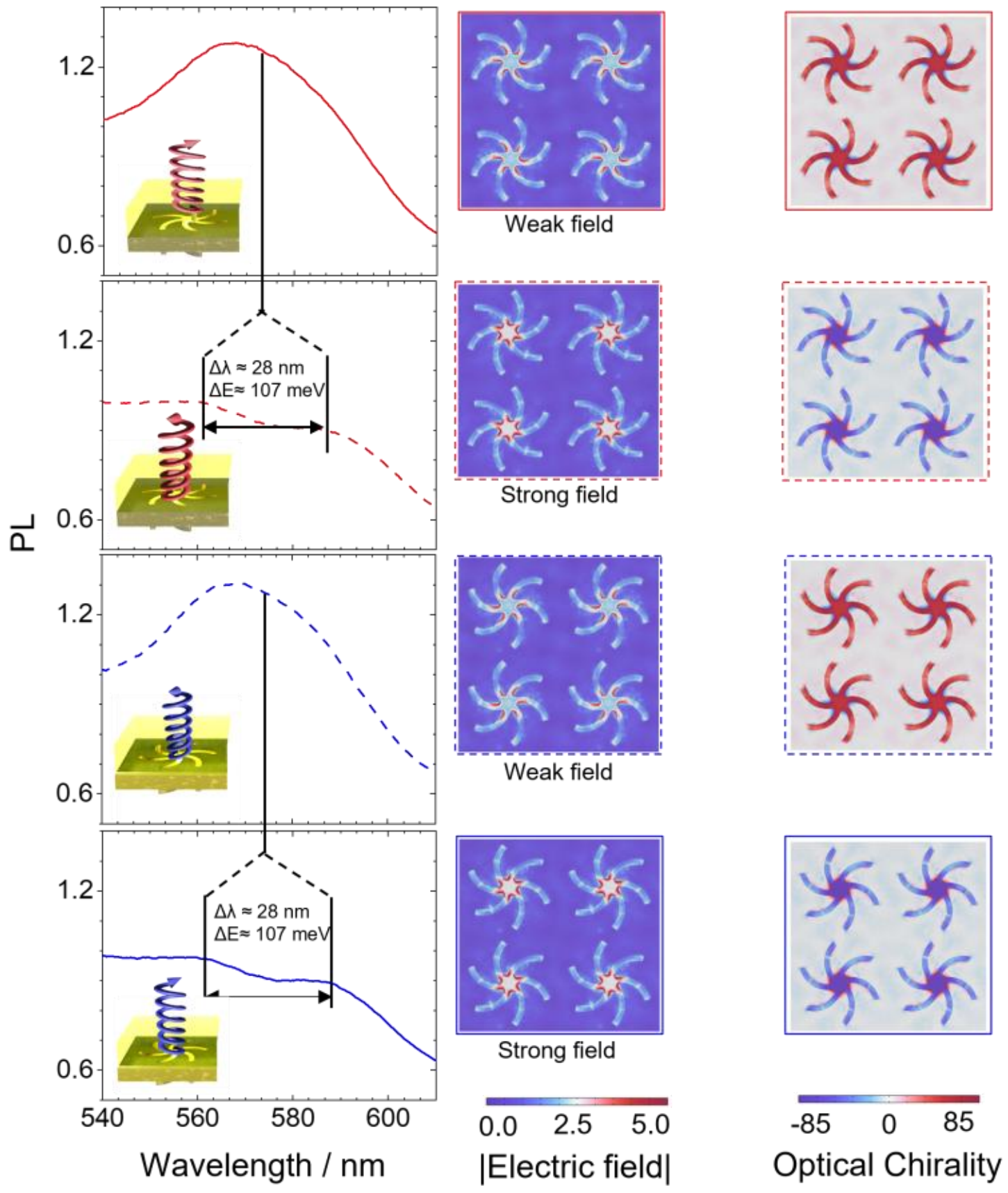

***Figure 4*** *The zoomed version of PL (left), simulated relative electric-field intensity (middle) and Optical Chirality (right) for 50% concentration dye on 1000 nm metafilm. The field plots are at the centre of the cavity and have been normalised with corresponding values of the flat gold. The strong field confinement is observed for matched combination of metafilm handedness and input light helicity i.e., LH/LCP and RH/RCP, thus leading to splitting in PL spectrum as shown on the left.*

A comparison between dichroic spectra derived from the scattering and luminescence of circularly

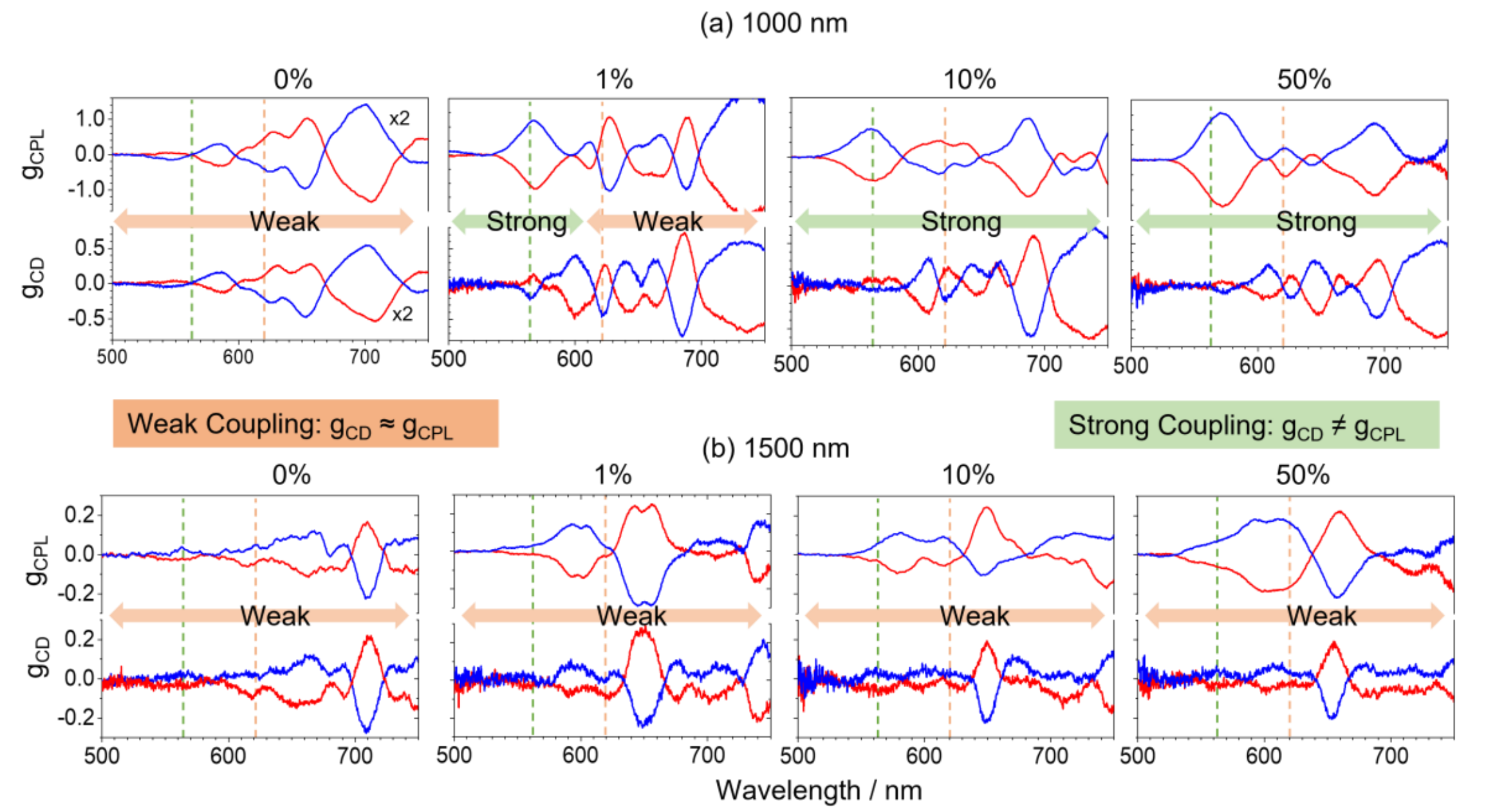

*Figure 5 (a) A comparison of CD ($g_{CD}$) and CPL($g_{CPL}$) spectra for (a) 1000 and (b) 1500 nm.* $g_{CD/CPL} = 2\frac{I_{LCP}-I_{RCP}}{I_{LCP}+I_{RCP}}$ *where $I_{LCP}$ and $I_{RCP}$ are the intensities of reflected / emitted LCP and RCP light. The concentration of dye for given column is shown on the top. The qualitative similarity of the spectra is indicated along with the predominate nature of the strength of coupling. The positions of the X and L inter-band transitions are labelled with green and orange dashed line respectively.*

polarised light, figure 5, which in accordance with literature will be referred to as CD and CPL spectra respectively, provides a means from discriminating between weak and stronger forms of coupling. In the weak coupling regime, such as that observed in the present work in the absence of the emitter, and in previous studies of emitter and chiral nanostructures [11, 12] the is a direct correlation between the CPL and CD spectra in terms of the position and sign of resonances. This is because under the weak coupling regime both the CD and CPL responses are governed by the *C* of the near field environments[33]. For the 1500 nm there is a correlation between CPL and CD spectra for all emitter doping, in agreement with the assignment of weak coupling. In contrast, there is a clear disparity between the CPL and CP spectra in the region covered by the L and X inter-band transitions at all dopant levels, consistent with stronger coupling. At longer wavelength, the region where plasmonic modes dominate, the level of correlation between CD and CPL is dependent on dopant concentration. For the 1% there is a correlation between CD and CPL, which is absence for higher dopant levels. The dependence of strong coupling on emitter concentration is well established[34]. CPL and CD data, which display similar qualitative behaviour, obtained for Rhodamine 6G in PMMA on related chiral nanocavities is displayed in supplementary information. This points to the coupling beyond the weak regime is not molecule specific.

Thus, using a luminescence spectral signature, band splitting, of strong coupling, supported by numerical simulations we demonstrate the control of strong coupling based on the symmetry combination of nanocavity handedness of light helicity. Strong coupling is achieved for Left Nanocavity / LCP and Right Nanocavity / RCP combinations, and therefore these can be considered to be left and right EM-enantiomers respectively.

# Conclusions

We have demonstrated symmetry control of the strength of chiral light - matter interactions between a geometrically achiral emitter molecule in a chiral nanocavity. The molecule displays distinctly different luminescence behaviour under strong, peak splitting, and weak, enhanced emission through the Purcell effect, coupling. This provides a simple route to create efficient (~30%) circularly polarised emission. Such a level of efficiency has previously required structurally complex and multicomponent chemical systems. A potential application of the phenomenon is in making the manufacture of OLED sources of CP light a less challenging chemical problem, since current strategies rely on the challenging synthesis and isolation of chiral geometric molecular structuresin high enantiomeric excess.

The results of this study and the concept of creating EM-enantiomers through strong coupling to chiral near fields has broader implications in chemical science. The current paradigm of polaritonic chemistry is based on controlling chemical reactivity in optical cavities through manipulating the energy landscape by strong coupling. Intriguingly, strong coupling occurs even in the absence of incident light through vacuum fluctuations, which is the zero-point energy of the quantised EM field inside an optical cavity. Consequently, just placing molecules in an optical cavity alters chemical reactivity. We propose a previously unconsidered concept that chiral induction in initially achiral molecules can be achieved by placing them in an appropriate chiral photonic cavity. This concept of the EM-enantiomer presages a straightforward route for instilling chirality into chemical systems, and hence simplify the synthesis of chiral materials.

# Acknowledgement

Affar Karimullah would like to acknowledge support by the UKRI, EPSRC (EP/S001514/1 and EP/S029168/1) and the James Watt Nanofabrication Centre. Shun Hashiyada acknowledges support from RIKEN Special Postdoctoral Researcher Program and JSPS (JP19K15514 and JP21K14594). Anastasiia Klimash thanks the EPSRC for funding (EP/T013710/1). Nikolaj Gadegaard and Malcolm Kadodwala acknowledge support from EPSRC (EP/S012745/1 and EP/S029168/1). Malcolm Kadodwala also acknowledges the Leverhulme Trust (RF-2019-023).